# The surface termination of a Fe (III) spin crossover molecular salt

M. Zaid Zaz[1], Binny Tamang[2], Kayleigh McElven[2], Esha Mishra[1,3], Gauthami Viswan[1], Wai Kiat Chin[1], Arjun Subedi[1], Alpha T. N'Daiye[4], Rebecca Y. Lai[2], and Peter A. Dowben[1]

[1] Department of Physics and Astronomy, Jorgensen Hall, University of Nebraska-Lincoln, Lincoln, NE 68588-0299, USA.

[2] Department of Chemistry, Hamilton Hall, University of Nebraska-Lincoln, Lincoln, NE, 68588-0304, USA.

[3] Department of Physics, Berry College, 2277 Martha Berry Hwy. NW., Mount Berry, GA 30149, USA.

[4] Advanced Light Source, Lawrence Berkeley National Laboratory, Berkeley, CA 94720, USA.

E-mail: zzaz2@huskers.unl.edu, pdowben@unl.edu

**Abstract**

From a comparison of energy dispersive analyses of X-rays (EDAX), the known molecular stoichiometry and X-ray photoemission spectroscopy (XPS), it is evident that the Fe(III) spin crossover salt [Fe(qsal)$_2$Ni(dmit)$_2$] has a preferential surface termination with the Ni(dmit)$_2$ moiety. This preferential surface termination leads to a significant surface to bulk core level shift for the Ni 2p X-ray photoemission core level, not seen in the corresponding Fe 2p core level spectra. Inverse photoemission spectroscopy (IPES), thus provides some indication of the density of states resulting from the dmit$^{2-}$ = 1,3-dithiol-2-thione-4,5-dithiolato ligand unoccupied molecular orbitals.

Keywords: Spin crossover molecules, surface to bulk core level shift, surface termination, X-ray absorption spectroscopy

## 1. Introduction

Molecular spin crossover systems are typically 3d transition metal complexes that undergo a reversible transition between a diamagnetic low spin state and a paramagnetic high spin state [1]. Although molecular systems, it has been increasingly well accepted that the surface can affect the spin state transition [2–7] so that the high spin state is populated at much lower temperatures at the surface than is the case for the bulk. This preference for the high spin state on the surface has now been seen in a variety of Fe (II) mononuclear spin crossover complexes [2,5,7], spin crossover polymers [3,4], and dinuclear spin crossover complexes [5,6]. While for the dinuclear species [Fe(3-Fpy)$_2${Ni(CN)$_4$}] [5], [Fe(3-Fpy)$_2${Pt(CN)$_4$}] [5] and {Fe(pz)[Pt(CN)4]} [6] it was conclusively demonstrated that the spin state transition at the surface differed from the bulk, little effort was made to investigate the surface termination. This is important because in such dinuclear spin crossover molecular salts, a preferential surface termination is possible, i.e. the preferential placement of one moiety over another may lower the total free energy of the system. This in turn could lead to a surface to bulk core level shift in X-ray photoemission [8–11], in spite of weak interactions expected of a molecular system, simply because the more limited coordination at the surface could reduce screening of the core hole, in the photoemission final state, compared to the bulk.

To explore the possibility that a preferential surface termination could exist and lead to a surface to bulk core level shift in the X-ray photoemission core level spectra, in this study we investigated on the surface termination of [Fe(qsal)$_2$Ni(dmit)$_2$], an Fe (III) SCO salt, where qsal = N(8quinolyl)salicylaldimine, and dmit$^{2-}$ = 1,3-dithiol-2-thione-4,5-dithiolato. This dinuclear spin crossover molecular complex adopts a layered structure [12], consisting of alternating Fe(qsal)$_2$ cationic and Ni(dmit)$_2$ anionic layers as seen in the inset to Figure 1, and is particularly appealing because the related Fe$^{III}$(qsal)$_2$Ni(dmit)$_2$]$_3$·CH$_3$CN·H$_2$O has a resistance less than 1 Ω.cm [13].

## 2. Experimental

The [Fe(qsal)$_2$Ni(dmit)$_2$] complex was synthesized as described in [12]. This Fe (III) complex features a gradual spin state transition as evidenced by the magnetic susceptibility times temperature *versus* temperature measurements (supplementary material), which matches well with what has been previously reported [12]. The XPS spectra were obtained at the Ni and Fe 2p core level using an Al k-alpha X-ray source, emitting X-rays at 1486.7 eV and hemispherical a SPECS PHOIBOS150 energy analyzer with a pass energy of 20 eV. A standard Shirely-Touggard type background [14–16] was applied to the raw spectra. The IPES spectra were obtained using a using an electron gun (Kimbell Physics), with electron kinetic energies within the range of 4-18 eV, combined with a channeltron detector (OmniVac). The X-ray absorption spectroscopy (XAS) spectra were obtained at beamline 6.3.1 at the Advanced Light Source at Lawrence Berkeley National Laboratories. The bulk composition of [Fe(qsal)$_2$Ni(dmit)$_2$] was characterized energy dispersive analysis of X-rays (EDAX) using an energy dispersive spectrometer (EDS), in a FEI Helios NanoLab 660 scanning electron microscope, at an incident electron energy of 20 keV. The EDXS acquisition time for the spectra was 300 seconds, with a scanning region of approximately 1600 µm$^2$.

The room temperature DC current-voltage I(V) measurements, of [Fe(qsal)$_2$Ni(dmit)$_2$] drop-cast on gold interdigitated electrodes (Au-IDE), was taken using a two-point probe method and recorded with a 4200A SCS parameter analyzer connected to a Lakeshore cryogenic probe station. The measurements were carried out in the absence of illumination.

## 3. Results and Discussion

The room temperature Fe 2p$_{3/2}$ XPS core level line shape, of [Fe(qsal)$_2$Ni(dmit)$_2$] as shown in Figure 1, is consistent with the high the high spin state [4–6,17–23] and thus consistent with the magnetometry [12]. The broadening of the Fe 2p$_{3/2}$ XPS core level feature, in particular, is caused by the final state effects in the form multiplets resulting from the interaction of the outgoing photoelectron with the unpaired electrons in the 3d orbitals [3-6,17]. Moreover, the metal composition ratio, derived from the Fe 2p (Figure 1) and Ni 2p (Figure 2) XPS core level spectra, after correcting for analyzer transmission function and photoemission cross-sections provides an Fe to Ni ratio of 1:2. This differs substantially from the nominal stoichiometry of [Fe(qsal)$_2$Ni(dmit)$_2$] and EDAX. EDAX reveals that Fe and Ni are present in almost 1:1 stoichiometric ratio, consistent with more bulk like probing depth of this spectroscopy and expected as per the synthesis protocol [12]. XPS is far more surface sensitive [8,24] and thus has strong surface contributions. So, the Fe to Ni ratio of 1:2, derived from the XPS Fe 2p and Ni 2p core level intensities, indicates surface segregation of the Ni(dmit)$_2$ moiety or more simply that the surface terminates in the Ni(dmit)$_2$ moiety.

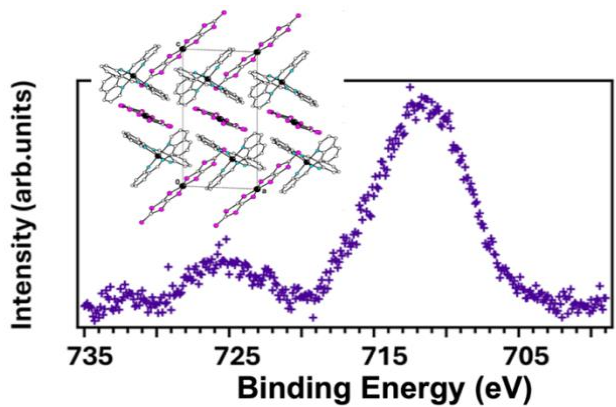

**Figure 1.** The *[Fe(qsal)$_2$Ni(dmit)$_2$] Fe 2p core level X-ray photoemission spectrum. The inset shows the layered crystal structure determined in [12]. The inset is reproduced from [12], C. Faulmann, et al., Inorganica Chimica Acta 360, 3870–3878 (2007), with permission from Elsevier.*

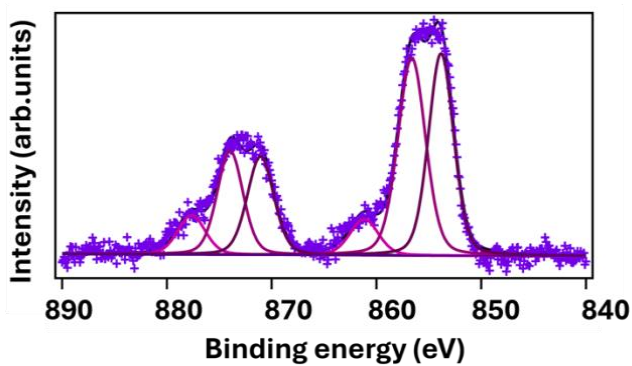

**Figure 2.** *The Ni 2p core level X-ray photoemission spectrum for [Fe(qsal)$_2$Ni(dmit)$_2$]. The fittings to the spectrum indicate the surface to bulk core level shift for the Ni(dmit)$_2$ moiety*

Consistent with the Ni(dmit)$_2$ moiety surface termination, XPS spectra obtained at the Ni 2p core level indicates a surface to bulk core level shift [8-11], applicable to the Ni(dmit)$_2$ moiety. In Figure 2, the Ni 2p$_{3/2}$ XPS core level envelope contains three main components, a low binding energy bulk core level at a binding energy of 853.8$\pm$0.2 eV, consistent with a bulk-like, or more highly coordinated Ni(dmit)$_2$ moiety, higher binding energy component at 856.7$\pm$0.2 eV, indicative of a surface Ni(dmit)$_2$ moiety and a two-hole bound state satellite at 861.2$\pm$0.2 eV binding energy. Combined with the very different Fe to Ni composition ratios obtained from XPS and EDAX, this is very compelling evidence for a Ni(dmit)$_2$ moiety surface termination. There is no known prior example of a surface to bulk core level shift, in the XPS core level binding energies of a spin crossover molecular complex [17], and a presence of a surface to bulk core level shift for [Fe(qsal)$_2$Ni(dmit)$_2$] is no doubt aided by the layered solid state structure for this complex. A surface to bulk core level shift as large as 2.9 eV, seen here, is actually quite significant and can be explained as a final state effect. [Fe(qsal)$_2$Ni(dmit)$_2$] is quite conducting and in fact the conductance is Ohmic, as seen in Figure 3 so that for Ni(dmit)$_2$, in the bulk, the photoemission final state would be well screened whereas the surface would not [25]. Contribution to the surface to bulk XPS core level shift from initial state effects cannot be excluded from the data here, but no such initial state effects appear in the Ni 2p X-ray absorption spectrum, as seen in Figure 4.

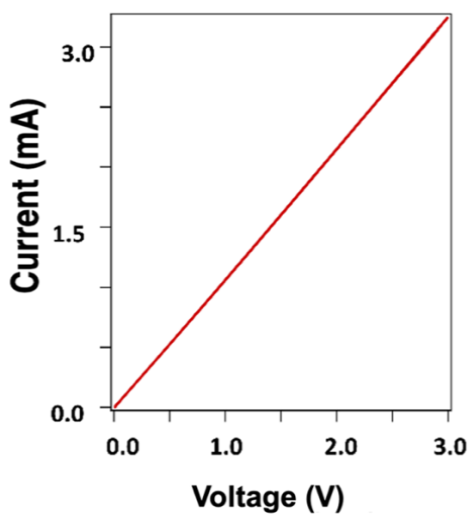

*Figure 3. The current versus voltage curve for [Fe(qsal)$_2$Ni(dmit)$_2$].*

If the surface terminates in the Ni(dmit)$_2$ moiety, then inverse photoemission should be particularly sensitive to the unoccupied molecular orbitals of the dmit ligand. To clarify this issue, we employed the more surface sensitive complementary IPES and XAS measurements. Owing to the extreme surface sensitivity [26,27], IPES spectral features will characteristically correspond to vacuum facing ligand weighted states, while the X-ray absorption (XAS) features correspond to metal weighted states [28]. Uniquely, the Ni 2p core level XAS spectra for [Fe(qsal)$_2$Ni(dmit)$_2$] show some spectral density well above the 2p$_{3/2}$ (L3) core threshold, as seen in Figure 4, not evident in the Fe 2p$_{3/2}$ XAS core level spectra. The IPES spectrum of [Fe(qsal)$_2$Ni(dmit)$_2$], presents a weak feature at about 3 eV above the Fermi level which lines up with the prominent metal feature in Ni L3 edge XAS spectrum. Also, the weak ligand features in the Ni XAS spectrum line up with prominent ligand features in IPES spectrum between 7.5 to 12.5 eV above the Fermi level. This alignment of the IPES spectrum with the Ni XAS spectrum is consistent with the notion that the surface of [Fe(qsal)$_2$Ni(dmit)$_2$], crystallites terminate in the Ni(dmit)$_2$ moiety.

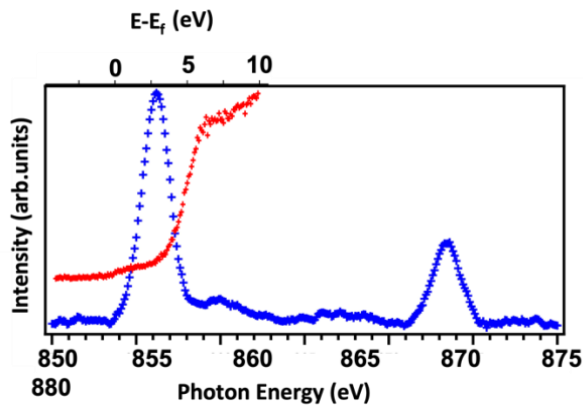

*Figure 4. The X-ray absorption spectra of [Fe(qsal)$_2$Ni(dmit)$_2$] obtained at the Ni 2p core, L3-L2, edge (Blue), inverse photoemission spectra of [Fe(qsal)$_2$Ni(dmit)$_2$] (red). The spectra are over laid to show the alignment of the dmit ligand features, prominent in IPES, with the high photon energy features in the XAS.*

## Conclusion

We have shown that the surface of the di-nuclear spin crossover molecular complex [Fe(qsal)$_2$Ni(dmit)$_2$], terminates in Ni(dmit)$_2$ moiety. As a consequence of this preferential termination, a surface to bulk core level shift in the Ni 2p XPS spectrum is observed. We thus provide conclusive proof that surface termination/segregation, which is a process driven by lowering of the total free energy is at play even in molecular crystallites. Additional evidence in favor of surface termination is provided in the form of complementary XAS and IPES measurements wherein, metal and ligand features from the Ni L3 X-ray absorption edge line up with the metal and ligand features observed in the IPES spectrum.

## Acknowledgements

This work was supported by the National Science Foundation (NSF) through the NSF-DMR-EPM 2317464 (MZZ, WKC, GV, EM and PAD), and EPSCoR RII Track-1: Emergent Quantum Materials and Technologies (EQUATE) (KAM, RL and BT). This research used resources of the Advanced Light Source, which is a DOE Office of Science User Facility under contract no. DE-AC02-05CH11231. MZZ acknowledges fruitful discussions with Archit Dhingra.